%% file: 0_main.tex
\documentclass[sigconf,natbib=true]{acmart}

\AtBeginDocument{%
  \providecommand\BibTeX{{%
    \normalfont B\kern-0.5em{\scshape i\kern-0.25em b}\kern-0.8em\TeX}}}

\usepackage{float}

\usepackage{booktabs}
\usepackage{array}
\usepackage{balance} 
\usepackage{multirow}
\usepackage[normalem]{ulem}
\usepackage{color}
\definecolor{lightgray}{RGB}{215,215,215}
\usepackage{colortbl}  
\usepackage{xcolor}
\useunder{\uline}{\ul}{}
\usepackage{subfigure}
\usepackage{algorithm}  
\usepackage{algorithmicx}  
\usepackage[noend]{algpseudocode}  

\usepackage{amsmath}  
\usepackage{enumitem}
\usepackage{tabularx}
\usepackage[utf8]{inputenc}
\usepackage[english]{babel}
\usepackage{amsthm}
\usepackage{bm}
\newcommand{\ie}{\emph{i.e., }}
\newcommand{\eg}{\emph{e.g., }}

\newcommand{\wrt}{\emph{w.r.t. }}
\newcommand{\cf}{\emph{cf. }}

\newlength\myindent
\floatname{algorithm}{Algorithm}

\copyrightyear{2024}
\acmYear{2024}
\setcopyright{acmlicensed}\acmConference[SIGIR '24]{Proceedings of the 47th International ACM SIGIR Conference on Research and Development in Information Retrieval}{July 14--18, 2024}{Washington, DC, USA}
\acmBooktitle{Proceedings of the 47th International ACM SIGIR Conference on Research and Development in Information Retrieval (SIGIR '24), July 14--18, 2024, Washington, DC, USA}
\acmDOI{10.1145/3626772.3657807}
\acmISBN{979-8-4007-0431-4/24/07}

\begin{document}

\title{Data-efficient Fine-tuning for LLM-based Recommendation}


\author{Xinyu Lin}
\email{xylin1028@gmail.com}
\affiliation{
\institution{National University of Singapore}
\city{}
\country{Singapore}
}
\author{Wenjie Wang$^*$}
\email{wenjiewang96@gmail.com}
\affiliation{
\institution{National University of Singapore}
\country{Singapore}
}
\author{Yongqi Li}
\email{liyongqi0@gmail.com}
\affiliation{
\institution{The Hong Kong Polytechnic University}
\city{Hong Kong SAR}
\country{China}
}
\author{Shuo Yang}
\email{syang98@hku.hk}
\affiliation{
\institution{The University of Hong Kong}
\city{Hong Kong SAR}
\country{China}
}

\author{Fuli Feng}
\email{fulifeng93@gmail.com}
\authornote{Corresponding author. This work is supported by the CCCD Key Lab of Ministry of Culture and Tourism.}
\affiliation{
\institution{University of Science and Technology of China}
\city{Hefei}
\country{China}
}
\author{Yinwei Wei}
\email{weiyinwei@hotmail.com}
\affiliation{
\institution{Monash University}
\city{Melbourne}
\country{Australia}
}
\author{Tat-Seng Chua}
\email{dcscts@nus.edu.sg}
\affiliation{
\institution{National University of Singapore}
\city{}
\country{Singapore}
}

\renewcommand{\shortauthors}{Xinyu Lin et al.}

\begin{abstract}

Leveraging Large Language Models (LLMs) for recommendation has recently garnered considerable attention, where fine-tuning plays a key role in LLMs' adaptation. However, the cost of fine-tuning LLMs on rapidly expanding recommendation data limits their practical application. To address this challenge, few-shot fine-tuning offers a promising approach to quickly adapt LLMs to new recommendation data. We propose the task of data pruning for efficient LLM-based recommendation, aimed at identifying representative samples tailored for LLMs' few-shot fine-tuning. 
While coreset selection is closely related to the proposed task, existing coreset selection methods often rely on suboptimal heuristic metrics or entail costly optimization on large-scale recommendation data.

To tackle these issues, we introduce two primary objectives for the data pruning task in the context of LLM-based recommendation:
1) high accuracy aims to identify the influential samples that can lead to high overall performance; and 
2) high efficiency underlines the low costs of the data pruning process. 
To pursue the two objectives, we propose a novel data pruning method incorporating two scores, namely influence score and effort score, to efficiently identify the influential samples. 
Particularly, the influence score is introduced to accurately estimate the influence of removing each sample on the overall performance. 
To achieve low costs of the data pruning process, we employ a small-sized surrogate model to replace LLMs to obtain the influence score. 
Considering the potential gap between the surrogate model and LLMs, we further propose an effort score to prioritize some hard samples specifically for LLMs. 
We instantiate the proposed method on two competitive LLM-based recommender models, and empirical results on three real-world datasets validate the effectiveness of our proposed method. 
In particular, our method uses only 2\% samples to surpass the full data fine-tuning, reducing time costs by 97\%.

\end{abstract}


\maketitle

\input{1_intro}
\input{2_0_task}
\input{2_1_method}

\input{3_exp}
\input{4_related_work}
\input{5_conclusion}

\clearpage

{
\tiny
\bibliographystyle{ACM-Reference-Format}
\balance
\bibliography{bibfile}
}

\newpage

\end{document}

%% file: 1_intro.tex
\section{Introduction}\label{sec:introduction}


Leveraging Large Language Models (LLMs) for recommendation has demonstrated promising efficacy across various tasks, including Click-Through Rate (CTR) prediction~\cite{bao2023tallrec}, sequential recommendation~\cite{rajput2023recommender}, and explainable recommendation~\cite{gao2023chat}. 
To build LLM-based recommender models, it is crucial to fine-tune LLMs on recommendation data for two primary reasons: 1) there exists a significant gap between previous LLMs' tuning tasks and the recommendation tasks~\cite{bao2023tallrec}, and 2) the rapid and continuous update of recommendation data necessitates frequent fine-tuning of LLMs~\cite{sachdeva2022infinite}. 
For example, there are approximately 160 million new videos and 942 billion interactions emerging on TikTok per day\footnote{https://www.tiktok.com/transparency/.}. 
Thus, frequent fine-tuning is imperative to incorporate up-to-date item information and enhance user behavior comprehension. 
However, fine-tuning LLMs on large-scale recommendation data demands substantial computational resources and time costs~\cite{li2023prompt}, 
thereby diminishing the practicality of LLM-based recommender models in real-world applications. 
As such, it is essential to enhance the fine-tuning efficiency of LLM-based recommender models. 

\begin{figure}[t]
\setlength{\abovecaptionskip}{0.05cm}
\setlength{\belowcaptionskip}{-0.3cm}
\centering
\includegraphics[scale=1.14]{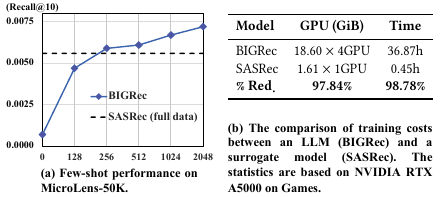}
\caption{(a) reveals that BIGRec achieves remarkable performance with only hundreds of samples. (b) shows the low costs of surrogate models. }
\label{fig:intro}
\end{figure}

Fortunately, the rich world knowledge encoded in LLMs offers a promising solution for efficient fine-tuning: \textit{few-shot fine-tuning}. 
Previous studies have uncovered that LLMs have the potential to quickly adapt to recommendation tasks 
by fine-tuning on randomly sampled few-shot data~\cite{bao2023tallrec,bao2023bi,lin2023multi} (Figure~\ref{fig:intro}(a)), significantly reducing training time and computational costs. 
Despite its efficiency, randomly sampled data may lack sufficient representativeness to enable LLMs to effectively comprehend new items and user behaviors. 
To combat this issue, we introduce the task of \textit{{data pruning for efficient LLM-based recommendation}}, 
which aims to identify representative samples tailored for LLMs' few-shot fine-tuning. 


A closely related literature to this data pruning task is coreset selection~\cite{guo2022deepcore}. 
It tries to select a small but representative subset from the full data, aiming to achieve comparable performance. 
Existing coreset selection methods generally fall into two categories\footnote{More detailed related work is discussed and compared in Section~\ref{sec:experiment} and~\ref{sec:related_work}.}: 
1) Heuristic methods select hard or diverse samples based on pre-defined metrics~\cite{paul2021deep,wu2023leveraging,luo2023recranker}. 
Such heuristic methods do not estimate the impact of selected samples on empirical risk, possibly leading to suboptimal coreset selection. 
2) Optimization-based methods mainly optimize the selection of subsets to minimize the empirical risk~\cite{borsos2020coresets,yang2022dataset}. 
However, these methods are inapplicable to large-scale recommendation datasets due to the complex and costly bi-level or discrete optimization problem~\cite{he2023large}. 
Worse still, both heuristic and optimization-based methods rely on the model well-trained by the full data to select the coreset, 
\eg calculating pre-defined scores or optimizing the data subset based on the well-trained model (\cf Section~\ref{sec:task_formulation}). 
As such, it is infeasible to directly apply these methods for LLM-based recommendation because of the high training costs of LLMs on the large-scale full recommendation data.

To overcome the above issues, we summarize two principal objectives for data pruning in the context of LLM-based recommendation: 
1) high accuracy, which focuses on selecting the samples that can lead to low empirical risk; and 
2) high efficiency, which emphasizes the low costs of the data pruning process, 
\ie eliminating the dependency of well-trained LLMs on the full data. 
Nevertheless, pursuing the two objectives faces two challenges: 
\begin{itemize}[leftmargin=*]
    \item To achieve high accuracy, it is essential to measure the influence of removing each training sample on the empirical risk. 
    However, assessing the influence of all samples is costly, as it requires the leaving-one-out retraining for each sample~\cite{tan2023data}. 
    \item To achieve high efficiency, one possible solution is to train a surrogate model for sample selection, \eg using a small-sized traditional recommender model, which can drastically reduce the GPU memory usage and the training time compared to LLMs (see Figure~\ref{fig:intro}(b)). 
    However, there exists a gap between LLMs and surrogate models, attributable to their divergent capabilities in learning user behaviors (refer to Figure~\ref{fig:learning_ability_gap}). 
    As such, influential samples selected by surrogate models might deviate from the ones on LLMs, potentially hurting the adaptation of LLMs. 
\end{itemize}

To address the challenges, we propose a novel \textbf{D}ata pruning method, to \textbf{E}fficiently identify the influenti\textbf{A}l samples for \textbf{L}LM-based \textbf{Rec}ommender fine-tuning (shorted as DEALRec). 
DEALRec leverages two scores, namely influence score and effort score, to identify the influential samples. 
The \textbf{\textit{influence score}} is formulated to estimate the influence of removing each training sample on the empirical risk. 
It is calculated by extending the influence function~\cite{hampel1974influence} via chain rules and second-order optimization techniques~\cite{koh2017understanding}. 
To efficiently calculate the influence score for all samples, 
DEALRec employs a simple yet effective symmetric property to accelerate the calculation, 
requiring only the estimation \textit{once} for all samples (\cf Section~\ref{sec:influence_score}). 
Thereafter, DEALRec uses a traditional recommender model as a surrogate model to obtain the influence score and introduces the \textbf{\textit{effort score}} to mitigate the gap between the surrogate model and LLMs. 
The effort score is obtained by calculating the gradient norm of a sample loss \wrt the parameters of LLMs,  
intuitively measuring the effort of LLMs to fit a specific sample. 
By regularizing the influence score with the effort score, DEALRec identifies the influential samples that encompass both the representativeness of the full data and the significance to LLMs. 
We instantiate DEALRec on two LLM-based recommender models and conduct extensive experiments on three real-world datasets, validating the superiority of DEALRec in terms of both efficiency and accuracy. 
The code and datasets are available at~\url{https://github.com/Linxyhaha/DEALRec}. 

In summary, this work offers three major contributions: 
\begin{itemize}[leftmargin=*]
    \item We introduce a data pruning task to identify the influential samples tailored for efficient LLM-based recommender fine-tuning, 
    unlocking the remarkable potential of applying LLM-based recommender models to real-world platforms. 
    
    \item We propose a novel data pruning method to discover the influential samples for LLM-based recommendation, which effectively and efficiently assesses the influence of removing a sample on empirical risk. 
    \item We conduct extensive experiments on three real-world datasets, demonstrating the effectiveness of DEALRec in achieving both high efficiency and accuracy. 
\end{itemize}

%% file: 2_0_task.tex
\section{Task Formulation}\label{sec:task_formulation}
In this section, we first introduce LLM-based recommender models and uncover the challenge of real-world applicability. 
Thereafter, we formulate the task of data pruning for LLM-based recommendation and compare the related work on coreset selection. 

\vspace{3pt}
\noindent$\bullet\quad$\textbf{LLM-based recommender models.} 
To leverage the competent capabilities of LLMs, LLM-based recommendation typically utilize powerful LLMs directly as the recommender models. 
Since LLMs are not particularly trained on the recommendation data, fine-tuning is the necessary and key step for LLMs to learn the item knowledge and understand user behavior. 
Let $\mathcal{U}$ and $\mathcal{I}$ denote the sets of users and items, respectively. 
We present each training sample, \ie user sequence, as $s=(x, y)$, where $x=[i_1, i_2, \dots, i_{|x|}]$ is the user's historical interactions in chronological order, and $y$ is the next interacted item of the user\footnote{Our main focus lies in sequential recommendation, which holds notable practical significance by intricately considering the temporal aspect in real-world scenarios.}, 
where $\{i_1, \dots, i_{|x|}, y\}\subset \mathcal{I}$. 
Formally, given the user sequences of the training set $\mathcal{D}=\{s_u|u\in\mathcal{U}\}$, the target is to fine-tune an LLM for recommendation tasks. 
The learnable parameters ($\phi\in\Phi$) of an LLM is optimized 
by minimizing the negative log-likelihood of the next interacted item $y$ conditioned on input $x$: 
\begin{equation}\label{eqn:task_llm_loss}\small
    \mathop{\min}_{\phi\in\Phi} \{\mathcal{L}_{\phi}^{LLM}=-\sum_{t=1}^{|y|} \log P_{\phi}(y_t|y_{<t},x)\}, 
\end{equation}
where $y_t$ denotes the $t$-th token of $y$, and $y_{<t}$ represents the token sequence preceding $y_t$. 

While fine-tuning LLMs has demonstrated effectiveness in recommendation tasks~\cite{liu2023once}, its practical application is hindered by the high resource costs required by LLMs and the continuous influx of new recommendation data~\cite{sachdeva2022infinite}. 
Hence, it is essential to enhance the efficiency of LLM-based recommender fine-tuning.

\vspace{3pt}
\noindent$\bullet\quad$\textbf{Data pruning for efficient LLM-based recommendation.} 
To achieve efficient LLM-based recommendation, a promising approach is to reduce the costs by few-shot fine-tuning with randomly selected samples~\cite{bao2023tallrec}. 
Nevertheless, the random samples might lose some crucial information for LLMs to acquire the latest information on user behavior or items, \eg trending items. 
In this light, we introduce the task of data pruning for efficient LLM-based recommendation, 
which aims to identify a set of representative samples particularly for LLMs' few-shot fine-tuning. 
Formally, given all training samples $\mathcal{D}=\{s_u|u\in\mathcal{U}\}$, the target of data pruning is to select a subset $\mathcal{S}\subset \mathcal{D}$, such that the LLMs trained on the subset $\mathcal{S}$ can yield good performance on the testing set. 
The size of $\mathcal{S}$ is controlled by the given selection ratio $r$, \ie $|\mathcal{S}|=r|\mathcal{D}|$. 

\vspace{3pt}
\noindent$\bullet\quad$\textbf{Retrospect of coreset selection.} 
As the closely related work to this data pruning task, coreset selection methods generally fall into two groups: 
\begin{itemize}[leftmargin=*]
    \item [1)] \textit{Heuristic} methods~\cite{coleman2020selection,toneva2018empirical,feldman2020neural} typically design some heuristic strategies to select samples based on an empirical minimizer:  
\begin{equation}\small\label{eqn:heuristic_based}
    \begin{aligned}
        \mathcal{S}=H(\hat{\theta},\mathcal{D}), 
        \quad \text{s.t.} \quad \hat{\theta}= \mathop{\arg\min}_{\theta\in\Theta}
        \mathcal{L}(\theta, \mathcal{D}), 
    \end{aligned}
    \end{equation}
    where $\mathcal{L}(\cdot)$ is the loss function of the task, \eg image classification~\cite{he2016deep} or CTR prediction~\cite{guo2017deepfm}, and $H(\cdot)$ denotes the heuristic strategy such as selecting samples with larger prediction entropy~\cite{coleman2020selection}, or clustering the samples based on the sample representations~\cite{chai2023efficient}. 
    However, this group of methods designs the strategy $H(\cdot)$ intuitively and fails to explicitly consider the influence of a sample on the empirical risk. 
    This might lead to suboptimal selection, thereby declining the performance of the model trained by the selected subset. 
    \vspace{2pt}
    \item [2)] \textit{Optimization-based} methods~\cite{borsos2020coresets,killamsetty2021retrieve,killamsetty2021glister,wu2023dataset} mainly utilize bi-level optimization techniques to learn the best subset chosen for training: 
\begin{equation}\small\label{eqn:optimization_based}
        \mathcal{S^{*}} = \mathop{\arg\min}_{\mathcal{S}\subset\mathcal{D}} \mathcal{L}(\hat{\theta}, \mathcal{D}), 
        \quad \text{s.t.} \quad \hat{\theta}=\mathop{\arg\min}_{\theta\in\Theta} \mathcal{L}(\theta, \mathcal{S}). 
    \end{equation}
    Besides, there is also some work that employs discrete optimization problems based on the empirical minimizer $\hat{\theta}$ in Eq. (\ref{eqn:heuristic_based}). 
    Nevertheless, 
    they struggle to be applied to large-scale datasets \eg recommendation data, due to the complex solving of the optimization problem~\cite{he2023large}. 
\end{itemize}

\noindent Furthermore, as shown in Eq. (\ref{eqn:heuristic_based}-\ref{eqn:optimization_based}), previous coreset selection methods usually require the model to be trained over original training samples $\mathcal{D}$, which however is infeasible for LLM-based recommender models due to the continuous influx of data and the high resource costs of LLMs (\cf Section~\ref{sec:introduction}). 

\vspace{2pt}
\noindent$\bullet\quad$ Drawing upon the above insights, we consider two objectives for data pruning: 
1) \textbf{\textit{high accuracy}} emphasizes the low empirical risk of the model trained on the selected samples, and 
2) \textbf{\textit{high efficiency}} focuses on the low costs of the data pruning process, breaking free from the heavy fine-tuning of LLMs for data pruning. 

%% file: 2_1_method.tex
\section{DEALRec}\label{sec:method}

\begin{figure}[t]
\setlength{\abovecaptionskip}{0.05cm}
\setlength{\belowcaptionskip}{-0.1cm}
\centering
\includegraphics[scale=1.13]{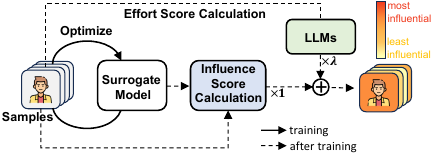}
\caption{Overview of DEALRec. DEALRec first trains a surrogate model on the full training samples. Subsequently, it calculates the influence score, which is then regularized by the effort score, to identify influential samples.}
\label{fig:method_overview}
\end{figure}

To pursue efficient LLM-based recommendation, we propose a novel data pruning method DEALRec, which involves two key components, \ie the \textit{influence score} to estimate the influence on empirical risk, and the \textit{effort score} as a regularization to mitigate the gap between surrogate model and LLMs. 
The overview of our method is presented in Figure~\ref{fig:method_overview}. 

\subsection{Influence Score}\label{sec:influence_score} 
To achieve good overall performance with the model trained on the pruned dataset $\mathcal{S}$, 
the key lies in the ability to assess the influence on the empirical risk, \ie overall performance, caused by removing a sample in training. 
However, simply assessing the the influence by removing each sample is impractical, because it requires brute force leaving-one-out-retraining for $n=|\mathcal{D}|$ times. 
To overcome this challenge, we propose an efficient approximation of the influence for all samples by extending influence on parameter change (\ie a classic result from influence function~\cite{koh2017understanding}) via chain rule and second-order optimization techniques. 
We further utilize the symmetric property to speed up the calculation of the influence score. 

\vspace{3pt}
\noindent$\bullet\quad$\textbf{Influence on parameter change.} 
To estimate the influence on empirical risk for each sample, we first start with the classic result~\cite{ling1984residuals} from research on influence function~\cite{cook1977detection}, which gives us the estimation of the parameter change caused by upweighting a sample $s$ for training. 
Considering a training sample $s$ is upweighted by a small $\epsilon$, the empirical minimizer can be rewritten as: 
\begin{equation}\label{eqn:reweight_minimizer}\small
    \hat{\theta}_{\epsilon, s}=\mathop{\arg\min}_{\theta\in\Theta} \frac{1}{n} \sum_{s_i\in\mathcal{D}}\mathcal{L}(s_i,\theta) + \epsilon\mathcal{L}(s,\theta).
\end{equation} 
According to~\cite{ling1984residuals}, the influence of upweighting a sample $s$ on the parameter change is then given as: 
\begin{equation}\label{eqn:influence_reweight}\small
    \mathcal{I}_{\text{param}}(s)=\frac{\mathrm{d}\hat{\theta}_{\epsilon,s}}{\mathrm{d}\epsilon}\bigg|_{\epsilon=0} = -H_{\hat{\theta}}^{-1}\nabla_{\theta}\mathcal{L}(s, \hat{\theta}),
\end{equation}
where $H_{\hat{\theta}} = \frac{1}{n}\sum_{s_i\in\mathcal{D}}\nabla_{\theta}^2\mathcal{L}(s_i, \hat{\theta})$ is the Hessian and positive definite by assumption, $\mathcal{I}_\text{param}(s)\in\mathbb{R}^{m}$, and $m$ is the number of parameters. 
Notably, assigning $-\frac{1}{n}$ to $\epsilon$ is equivalent to removing the sample $s$ from training. 
As such, the parameter change of removing a training sample $s$ can be linearly approximated as: 
\begin{equation}\label{eqn:influence_param_change}\small
    \hat{\theta}_{-s}-\hat{\theta}\approx - \frac{1}{n}\mathcal{I}_{\text{param}}(s)=\frac{1}{n}H_{\hat{\theta}}^{-1}\nabla_{\theta}\mathcal{L}(s,\hat{\theta}),  
\end{equation}
where $\hat{\theta}_{-s}=\mathop{\arg\min}_{\theta\in\Theta}\sum\nolimits_{s_i\in\mathcal{D}, s_i \ne s}\mathcal{L}(s_i,\theta)$.  

Based on Eq. (\ref{eqn:influence_param_change}), an intuitive approach to assess the sample influence for model training is to utilize the L2 norm of a sample's influence on parameter change or an additional discrete optimization problem as proposed in~\cite{yang2022dataset}. 
Nevertheless, large parameter changes do not necessarily lead to performance improvements. 
Besides, calculating Eq. (\ref{eqn:influence_param_change}) for all training samples can be computationally costly~\cite{he2023large} and is infeasible for recommendation data. 
To alleviate the issues, we propose an efficient approximation for the influence of removing a sample on the empirical risk. 

%

\vspace{3pt}
\noindent$\bullet\quad$\textbf{Influence on empirical risk.} 
Based on the parameter change obtained via the influence function, we can then estimate the influence of upweighting a training sample $s$ by a small $\epsilon$ on the loss of an arbitrary sample $s'$: 
\begin{equation}\small\label{eqn:influence_reweight_sample_loss}
\begin{aligned}
    \mathcal{I}_{\text{upweight,loss}}(s,s')&\overset{\text{def}}{=} \frac{\mathrm{d}\mathcal{L}(s', \hat{\theta}_{\epsilon,s})}{\mathrm{d}\epsilon}\bigg|_{\epsilon=0}  \\
    &=\nabla_{\theta}\mathcal{L}(s',\hat{\theta})^{\mathrm{T}}\frac{\mathrm{d}{\hat{\theta}_{\epsilon,s}}}{\mathrm{d}\epsilon}\bigg|_{\epsilon=0} \quad\quad \text{(chain rule)}\\
    &=-\nabla_{\theta}\mathcal{L}(s',\hat{\theta})^{\mathrm{T}} H_{\hat{\theta}}^{-1}\nabla_{\theta}\mathcal{L}(s,\hat{\theta}).
\end{aligned}
\end{equation}
Similarly, the influence of removing a training sample $s$ on the loss of an arbitrary sample $s'$ can be linearly approximated as: 
\begin{equation}\small\label{eqn:influence_remove_sample_loss}
    \mathcal{I}_{\text{remove, loss}}(s, s')=\frac{1}{n}\nabla_{\theta}\mathcal{L}(s',\hat{\theta})^{\mathrm{T}}H_{\hat{\theta}}^{-1}\nabla_{\theta}\mathcal{L}(s,\hat{\theta}).
\end{equation}
We can then obtain the influence of removing a sample $s$ on the empirical risk (\ie influence score) by 
\begin{equation}\small\label{eqn:influence_overall_inefficient}
\begin{aligned}
        \mathcal{I}_{\text{remove,loss}}(s,{\mathcal{D}})&{=} -\frac{1}{n}\frac{\mathrm{d}\left(\frac{1}{n}\sum\nolimits_i\mathcal{L}(s', \hat{\theta}_{\epsilon,s_i})\right)}{\mathrm{d}\epsilon}\Bigg|_{\epsilon=0}  \\
        &=\underbrace{\frac{1}{n}\sum\nolimits_i\frac{1}{n}\nabla_{\theta}\mathcal{L}(s_{i},\hat{\theta})^{\mathrm{T}}H_{\hat{\theta}}^{-1}\nabla_{\theta}\mathcal{L}(s,\hat{\theta})}_{\let\scriptstyle\textstyle\substack{(\textbf{influence score})}}. \\
\end{aligned}
\end{equation} 
However, it is non-trivial to directly obtain $H_{\hat{\theta}}^{-1}$ as forming and inverting $H_{\hat{\theta}} = \frac{1}{n}\sum_{s_i\in\mathcal{D}}\nabla_{\theta}^2\mathcal{L}(s_i, \hat{\theta})$ requires $\mathcal{O}(nm^2+m^3)$ with $n$ training samples and $\theta\in\mathbb{R}^{m}$. 
This results in cumbersome calculation of influence scores for all training samples. 

\begin{algorithm}[t]
    \small
    \caption{Procedure of HVP Estimation}  
    \label{algo:HVP_estimation}
    \begin{algorithmic}[1]
    \Require Original training dataset $\mathcal{D}$, parameters of a well-trained model $\hat{\theta}$, 
    iteration number $T$. 
    
    \State Compute $\sum\nolimits_i\frac{1}{n}\nabla_{\theta}\mathcal{L}(s_{i},\hat{\theta})$ for $\forall i \in \{1, \dots, n\}$. 
    \State Initialize 
    $\Tilde{H}_{0}^{-1} \left[\sum\nolimits_i\frac{1}{n}\nabla_{\theta}\mathcal{L}(s_{i},\hat{\theta})\right] = \sum\nolimits_i\frac{1}{n}\nabla_{\theta}\mathcal{L}(s_{i},\hat{\theta})$.
        
    \ForAll{$t\in\{1,\dots,T\}$}
        
        \State Randomly sample a training sample $s_t\in\mathcal{D}$; 
        \State Calculate $\nabla_{\theta}^2\mathcal{L}(s_{t})$ as the unbiased estimator of $H$; 
        \State
        $\Tilde{H}_t^{-1}\left[\sum\nolimits_i\frac{1}{n}\nabla_{\theta}\mathcal{L}(s_{i},\hat{\theta})\right] \leftarrow  \sum\nolimits_i\frac{1}{n}\nabla_{\theta}\mathcal{L}(s_{i},\hat{\theta}) + $
        \Statex $\quad \quad \quad \quad \quad \quad \quad \left[I-\nabla_{\theta}^2\mathcal{L}(s_{t})\right]\Tilde{H}^{-1}_{t-1} \left[\sum\nolimits_i\frac{1}{n}\nabla_{\theta}\mathcal{L}(s_{i},\hat{\theta})\right]$;
        \algorithmiccomment{Eq. (\ref{eqn:hvp_unbiased_update})}
        
    \EndFor
    \State $\Tilde{H}^{-1} \left[\sum\nolimits_i\frac{1}{n}\nabla_{\theta}\mathcal{L}(s_{i},\hat{\theta})\right] \leftarrow \Tilde{H}^{-1}_T \left[\sum\nolimits_i\frac{1}{n}\nabla_{\theta}\mathcal{L}(s_{i},\hat{\theta})\right]$.
    \Ensure Unbiased estimation $\Tilde{H}^{-1} \left[\sum\nolimits_i\frac{1}{n}\nabla_{\theta}\mathcal{L}(s_{i},\hat{\theta})\right]$.
	\end{algorithmic}
  
\end{algorithm}

\vspace{3pt}
\noindent$\bullet\quad$\textbf{Efficient estimation of influence score.} 
To achieve efficient computation of influence score, we utilize stochastic-based Hessian-Vector Products (HVP)~\cite{agarwal2016second} to efficiently approximate 
$H_{\hat{\theta}}^{-1} \nabla_{\theta}\mathcal{L}(s,\hat{\theta})$. 
The idea of stochastic-based HVP estimation is to iteratively obtain an unbiased estimator of $H_{\hat{\theta}}$ and approach the unbiased estimation of HVP, \ie $H_{\hat{\theta}}^{-1} \nabla_{\theta}\mathcal{L}(s,\hat{\theta})$. 
Specifically, we omit the $\hat{\theta}$ subscript for clarity and write the first $j$ terms in Taylor expansion of $H^{-1}$ as $H_j^{-1}\overset{\text{def}}{=}\sum_{i=0}^{j}(I-H)^{i}$, 
which can be further rewritten recursively as $H_j^{-1}=I+(I-H)H_{j-1}^{-1}$. 
From the validity of the Taylor expansion, we have $H_j^{-1}\rightarrow H^{-1}$ as $j\rightarrow \infty$. 
Thereafter, denoting $\nabla_{\theta}\mathcal{L}(s,\hat{\theta})$ as $v$, the update iteration for the estimated $H_{\hat{\theta}}^{-1} \nabla_{\theta}\mathcal{L}(s,\hat{\theta})$ at step $t$ can be written as: 
\begin{equation}\small\label{eqn:hvp_unbiased_update}
    \Tilde{H}_t^{-1}v=v+\left(I-\nabla_{\theta}^2\mathcal{L}(s_{t})\right)\Tilde{H}^{-1}_{t-1}v,
\end{equation}
where $s_t$ is a training sample randomly drawn from $\mathcal{D}$, and $\nabla_{\theta}^2\mathcal{L}(s_{t})$ is an unbiased estimator of the $H$ at step $t$ for fast-to-compute HVP~\cite{koh2017understanding}. 
\vspace{2pt}
Despite that stochastic-based HVP can alleviate the computation burdens of the estimation, calculating the influence score for each sample is still costly due to the independent $n$ estimations of $H_{\hat{\theta}}^{-1} \nabla_{\theta}\mathcal{L}(s,\hat{\theta})$ for each $s\in\mathcal{D}$ (refer to Eq. (\ref{eqn:influence_overall_inefficient})).

\vspace{2pt}
To further enhance the efficiency of acquiring influence scores for all samples, we use symmetric property to rewrite Eq. (\ref{eqn:influence_overall_inefficient}) into:
\begin{equation}\small\label{eqn:influence_overall_efficient}
\begin{aligned}
        \mathcal{I}_{\text{remove,loss}}(s,{\mathcal{D}})
        &=\frac{1}{n} \nabla_{\theta}\mathcal{L}(s,\hat{\theta})^{\mathrm{T}} \underbrace{H_{\hat{\theta}}^{-1} \left[\sum\nolimits_i\frac{1}{n}\nabla_{\theta}\mathcal{L}(s_{i},\hat{\theta})\right]}_{\let\scriptstyle\textstyle\substack{(\textbf{constant vector})}}.
\end{aligned}
\end{equation}
The reformulation is based on the assumption that $\mathcal{L}(\cdot)$ has continuous second-order derivatives, which is consistent with the assumption for influence function~\cite{koh2017understanding}, leading to the fact that $H^{-1}_{\hat{\theta}}$ is symmetric.  
Since $H_{\hat{\theta}}^{-1} \left[\sum\nolimits_i\frac{1}{n}\nabla_{\theta}\mathcal{L}(s_{i},\hat{\theta})\right]\in\mathbb{R}^{m}$ is a constant vector for any sample $s\in\mathcal{D}$, 
we can efficiently obtain influence scores for all samples by only applying HVP estimation \textbf{once} for $H_{\hat{\theta}}^{-1} \left[\sum\nolimits_i\frac{1}{n}\nabla_{\theta}\mathcal{L}(s_{i},\hat{\theta})\right]$. 
The detailed HVP estimation process is illustrated in Algorithm~\ref{algo:HVP_estimation}. 



\begin{figure}[t]
\setlength{\abovecaptionskip}{0.05cm}
\setlength{\belowcaptionskip}{-0.05cm}
\centering
\includegraphics[scale=1]{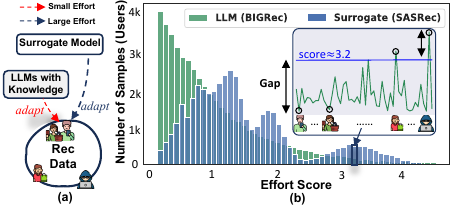}
\caption{{(a) depicts the different learning ability due to the prior knowledge in LLMs. 
(b) presents the distributions of effort scores of LLM and surrogate model on Games dataset\protect\footnotemark.
}}
\label{fig:learning_ability_gap}
\end{figure}
\footnotetext{We obtain the effort scores for surrogate model by calculating the gradient norm of the parameters of the surrogate model (Eq. (\ref{eqn:user_gap})).}

\subsection{Gap Regularization}\label{sec:gap_regularization}
As shown in Eq. (\ref{eqn:influence_overall_efficient}), assessing the influence score of a sample requires the optimized parameters $\hat{\theta}$ well-trained over all training samples $\mathcal{D}$. 
Nevertheless, this poses challenges for LLM-based recommender models due to the continuous influx of large-scale new data in real-world scenarios. 
In this light, we propose to utilize a surrogate model to replace the LLMs and introduce an \textbf{\textit{effort score}} as a gap regularization to complement the learning ability gap between LLMs and the surrogate models. 

\vspace{3pt}
\noindent$\bullet\quad$\textbf{Surrogate model.} 
To reduce the costs, we propose utilizing a surrogate model, \eg a small-sized traditional recommender model, to compute the influence scores. 
Nevertheless, since LLMs acquire rich world knowledge during the pre-training stage, they intricately possess different learning abilities compared to the surrogate model (Figure~\ref{fig:learning_ability_gap}(a)). 
Therefore, the influential samples on LLMs might deviate from the ones for LLMs. 

\vspace{3pt}
\noindent$\bullet\quad$\textbf{Effort score.}
To compensate for the gap, we introduce the effort score, 
which aims to capture significant samples particularly for LLMs. 
Specifically, we define the effort score of a sample, \ie a user sequence, $s$ as:
\begin{equation}\label{eqn:user_gap}
\delta_s=\lVert\nabla_{\phi}\mathcal{L}^{LLM}(s)\rVert_{2},
\end{equation}
where $\phi$ is the learnable parameters of LLMs\footnote{The learnable parameters can be either the whole parameters of LLMs or the learnable parameters from parameter-efficient training, \eg LoRA~\cite{hu2021lora}.}. 
Intuitively, it measures the learning effort of LLMs to fit a specific user sequence, and a larger score indicates a harder sample for LLMs to learn. 
To elaborate, Eq. (\ref{eqn:user_gap}) measures the change in the model parameters, which can be interpreted as the discrepancy from the current knowledge encoded in LLMs' parameters to the latest item knowledge or user behavior. 
As such, the effort score can emphasize significant samples particularly for LLMs, supplementing the different learning ability of the surrogate model (Figure~\ref{fig:learning_ability_gap}(b)). 



\begin{algorithm}[t]
    \small
    \caption{Procedure of DEALRec}  
    \label{algo:DEALRec}
    \begin{algorithmic}[1]
    \Require Original training dataset $\mathcal{D}$, randomly initialized parameters of surrogate model $\theta$, pre-trained parameters of LLM $\phi$. 

    \State $\Hat{\theta}=\mathop{\arg\min}_{\theta\in\Theta}\frac{1}{n}\sum\nolimits_{s_i\in\mathcal{D}}\mathcal{L}(s_i, \theta)$.
    \State Obtain estimated ${H}^{-1} \left[\sum\nolimits_i\frac{1}{n}\nabla_{\theta}\mathcal{L}(s_{i},\hat{\theta})\right]$ via HVP estimation. 
    \ForAll{$ i\in\{1,\dots, n\}$}
        \State $I_{s_i} = \frac{1}{n^2} \nabla_{\theta}\mathcal{L}(s_i,\hat{\theta})^{\mathrm{T}} H_{\hat{\theta}}^{-1} \left[\sum\nolimits_j\frac{1}{n}\nabla_{\theta}\mathcal{L}(s_{j},\hat{\theta})\right] + $ 
        \Statex $ \quad\quad \quad\quad\quad\quad \quad\quad\quad\quad \quad\quad\quad\quad \quad\quad \lambda \lVert\nabla_{\phi}\mathcal{L}^{LLM}(s_i)\rVert_{\scriptstyle{2}}$;
        \algorithmiccomment{Eq. (\ref{eqn:final_score})}
    \EndFor
    \State $\mathcal{G}=\{G_1,\dots,G_K\} \leftarrow$ Split training samples $\mathcal{D}$ into $K$ groups according to the final score $I_s$ with even range width.  
    \State $\mathcal{S}\leftarrow \varnothing$, $B\leftarrow \lfloor\frac{r|\mathcal{D}|}{K}\rfloor$.

    \While{$\mathcal{G} \ne \varnothing$}
        \State $k^*=\mathop{\arg\min}_{k}|G_k|$; 
        \State $\mathcal{S}_{k^*}\leftarrow$ randomly select $\mathop{\min}\{B,|G_{k^*}|\}$ samples from $G_{k^*}$;
        \State $\mathcal{S}\leftarrow \mathcal{S} \cup \mathcal{S}_{k^*}$; $\mathcal{G}\leftarrow \mathcal{G}\setminus \{G_{k^*}\}$;
        \State $B\leftarrow \lfloor\frac{r|\mathcal{D}|-|\mathcal{S}|}{|\mathcal{G}|}\rfloor$;
        \algorithmiccomment{Update sampling budget}
    \EndWhile
    
    \Ensure Selected samples $\mathcal{S}$ for few-shot fine-tuning.
	\end{algorithmic}
  
\end{algorithm}

\vspace{3pt}
\noindent$\bullet\quad$\textbf{Overall score.} 
By injecting the signals of LLMs' learning ability into the calculation of influence score, we can obtain the final score of each user sequence for LLM-based recommender fine-tuning: 
\begin{equation}\label{eqn:final_score}\small
    I_s = \underbrace{\frac{1}{n^2} \nabla_{\theta}\mathcal{L}(s,\hat{\theta})^{\mathrm{T}} H_{\hat{\theta}}^{-1} \left[\sum\nolimits_i\frac{1}{n}\nabla_{\theta}\mathcal{L}(s_{i},\hat{\theta})\right]}_{\let\scriptstyle\textstyle\substack{(\textbf{influence score})}} + \underbrace{\lambda \lVert\nabla_{\phi}\mathcal{L}^{LLM}(s)\rVert_{2}}_{\let\scriptstyle\textstyle\substack{(\textbf{effort score})}}, 
\end{equation}
where $\lambda$ is a hyper-parameter to balance the strength of the gap regularization. 
Notably, 
the gap regularization would suppress the easy samples with smaller effort scores while emphasizing the samples that are more difficult to learn, \ie larger effort scores. 

Intuitively, DEALRec identifies the influential samples with two key considerations: 1) the influence score focuses on selecting the representative samples from the full dataset, capturing collaborative filtering information for low empirical risk; and 2) the effort score highlights the non-trivial samples that are significant to the learning of LLMs. 
The effectiveness of the two scores is empirically validated in Section~\ref{sec:ablation}.


\subsection{Few-shot Fine-tuning}\label{sec:sample_selection}
Based on the final influential score obtained via Eq. (\ref{eqn:final_score}), we can select a subset of data $\mathcal{S}$ for LLMs' few-shot fine-tuning, given an expected selection ratio $r$.

\vspace{3pt}
\noindent$\bullet\quad$\textbf{Few-shot data coverage.} 
A straightforward approach is to select the data greedily, \ie rank the samples based on the overall scores, and then select the top-$r$ percentage of the training data. 
However, greedily selecting the samples with higher scores might result in very similar samples with low data coverage, which leads to: 
1) Inadequacy of samples from other areas, thus hurting the bounded empirical risk~\cite{zheng2022coverage} and lowering the overall performance (\cf Section~\ref{sec:overall_performance}). 
2) Poor utilization of training samples because of the redundant samples with similar patterns, thereby causing suboptimal selection for few-shot fine-tuning. 


\vspace{3pt}
\noindent$\bullet\quad$\textbf{Coverage-enhanced sample selection.} 
To address the above issues, we follow~\cite{zheng2022coverage} to select the users based on the idea of stratified sampling. 
The core idea is to maintain the budget for the samples in different areas of training distribution, such that the data coverage will be improved to ensure a high-probability bound for the empirical risk (refer to~\cite{zheng2022coverage} for detailed proof). 
In detail, we first divide the samples into $K$ groups according to their overall scores. 
We then iteratively sample $n_s$ user sequences from the group with the fewest samples and discard that group after sampling, where $n_s$ is the average sampling budget for all groups and is initialized with $\lfloor\frac{r|\mathcal{D}|}{K}\rfloor$. 
If the group size is smaller than the average sampling budget, we select all users from this group and update the average sampling budget for the remaining groups (see Algorithm~\ref{algo:DEALRec}). 

\vspace{2pt}
Based on the selected few-shot samples $\mathcal{S}$, we optimize the learnable parameters ($\phi\in\Phi$) of LLMs: 
\begin{equation}\label{eqn:fewshot_llm_optimization}\small
    \hat{\phi} = \mathop{\arg\min}_{\phi\in\Phi} \frac{1}{|\mathcal{S}|}\sum_{s_i\in\mathcal{S}}\mathcal{L}_{\phi}^{\text{LLM}}(s_i).  
\end{equation}

\vspace{3pt}
\noindent$\bullet\quad$\textbf{Instantiation.} 
To instantiate DEALRec on LLM-based recommender models, we first employ a surrogate model to train on original training samples $\mathcal{D}$ and calculate the influence score for all samples via Eq. (\ref{eqn:influence_overall_efficient}), where the $\mathcal{L}(\cdot)$ can be any form of the loss function from the surrogate model, \eg BPR~\cite{rendle2009bpr}. 
We then obtain the effort score for LLMs via Eq. (\ref{eqn:user_gap}), where $\phi$ can be the learnable parameters from any backend LLM-based recommender models. 
Eventually, we apply the stratified sampling to select the samples for LLMs' few-shot fine-tuning. 
The detailed data pruning process of DEALRec is demonstrated in Algorithm~\ref{algo:DEALRec}.

%% file: 3_exp.tex
\section{Experiment}\label{sec:experiment}

We conduct extensive experiments on three real-world datasets to answer the following research questions: 
\begin{itemize}[leftmargin=*]
    \item \textbf{RQ1:} How does our proposed DEALRec perform compared to the coreset selection baselines for LLM-based recommendation and the models trained with full data? 
    \item \textbf{RQ2:} How do the different components of DEALRec (\ie influence score, gap regularization, and stratified sampling) affect the performance, and is DEALRec generalizable to different surrogate models? 
    \item \textbf{RQ3:} How does DEALRec perform under different selection ratios and how does DEALRec improve the overall performance? 
\end{itemize}

\subsection{Experimental Settings}

\subsubsection{\textbf{Datasets.}}\label{subsubsec:datasets}

\begin{table}[t]
\setlength{\abovecaptionskip}{0.0cm}
\setlength{\belowcaptionskip}{0cm}
\caption{Statistics of the three datasets.}
\setlength{\tabcolsep}{2.2mm}{
\resizebox{0.46\textwidth}{!}{
\begin{tabular}{lcccc}
\toprule
\textbf{Datasets} & \textbf{\# Users} & \textbf{\# Items} & \textbf{\# Interactions} & \textbf{Density} \\ \midrule
\textbf{Games} & 49,156 & 17,332 & 342,329 & 0.04\% \\
\textbf{MicroLens-50K} & 49,887 & 19,217 & 359,048 & 0.04\% \\
\textbf{Book} & 88,263 & 86,272 & 5,303,707 & 0.07\% \\ \bottomrule
\end{tabular}
}}
\label{tab:dataset_statistics}
\end{table}

We conduct experiments on three real-world recommendation datasets: 
1) \textbf{Games} is from the Amazon review datasets\footnote{\url{https://jmcauley.ucsd.edu/data/amazon/.}}, which covers interactions between users and video games with rich textual features. 
2) \textbf{MicroLens-50K}\footnote{\url{https://github.com/westlake-repl/MicroLens/.}} is a newly released micro-video recommendation dataset~\cite{ni2023content}. It contains 50$k$ users' interactions with micro-videos and their associated multimodal features. 
3) \textbf{Book} is also from Amazon review datasets, containing users' interactions with extensive books. 
For Games and Book, we follow previous work and discard the interactions with the ratings $< 4$. For the three datasets, we sort all user-item interactions according to the global timestamps, and then split the interactions into training, validation, and testing sets with the ratio of 8:1:1. 
Besides, we consider two different fine-tuning settings as follows: 
1) \textit{Few-shot fine-tuning} fine-tunes LLM-based recommender models with limited samples at a fixed size, \eg 1024-shot, obtained via different data pruning methods. 
2) \textit{Full fine-tuning} utilizes all samples to fine-tune LLM-based recommender models without data pruning. 

\subsubsection{\textbf{Baselines.}}
We compare DEALRec with the random sampling and several competitive coreset selection methods, including difficulty-based methods and diversity-based methods: 
1) \textbf{Random} obtains the data subset via random sampling, which is a popular and strong baseline in data-efficient training~\cite{guo2022deepcore}. 
2) \textbf{GraNd}~\cite{paul2021deep} is a representative coreset selection method that selects the difficult samples with larger gradient norms during training. 
3) \textbf{EL2N}~\cite{paul2021deep} proposes to select the difficult samples with larger errors between the labels and the prediction from the model trained by the original dataset. 
4) \textbf{CCS}~\cite{zheng2022coverage} is a competitive method that selects the samples considering both high data coverage and sample importance. 
We use EL2N as the importance metric for CCS. 
5) \textbf{TF-DCon}~\cite{wu2023leveraging} is a recently proposed data pruning method for content-based recommendation, which clusters the user sequences based on the user representations obtained from both well-trained recommender models and LLMs for selection. 
6) \textbf{RecRanker}~\cite{luo2023recranker} proposes a sampling strategy to select high-quality user sequences. It selects the users with more interactions for better user modeling and utilizes a cluster-based sampling strategy to enhance user diversity. 

We do not perform optimization-based methods for comparison because of the inapplicability of complex bi-level or discrete optimization for LLMs on large-scale recommendation data (\cf Section~\ref{sec:task_formulation}). 
We instantiate our proposed DEALRec and all baselines on two competitive backend LLM-based recommender models: 
1) \textit{BIGRec}~\cite{bao2023bi} utilizes the item title to present the user sequence for recommendation generation; 
2) \textit{TIGER}~\cite{rajput2023recommender} learns extra tokens from item features to present items, and then converts the user sequence into the sequence of the new item token for next-item generation. 


\vspace{3pt}
\noindent$\bullet\quad$\textbf{Evaluation.}
We employ the widely used metrics Recall@$K$ and NDCG@$K$ to evaluate the models~\cite{he2020lightgcn}, 
with $K$ set to $10$ and $20$ for Games, and $K=20$ and $50$ for MicroLens-50K and Book\footnote{We report metrics@$20$ and @$50$ because of the challenging modeling of user behavior on book and micro-video recommendations, where the temporal shifts of user interests and the item feature is stronger and thus more difficult to capture~\cite{wang2023causal,wang2023equivariant}.}. 

\begin{table*}[t]
\setlength{\abovecaptionskip}{0.05cm}
\setlength{\belowcaptionskip}{0.2cm}
\caption{Overall performance comparison between the baselines and DEALRec instantiated on two competitive LLM-based recommender models on three datasets. For each backend model, the bold results highlight the best results while the second-best ones are underlined. $*$ implies the improvements over the second-best results are statistically significant ($p$-value < 0.01) under one-sample t-tests. We run all experiments for 3 times with different random seeds and report the averaged results. }
\setlength{\tabcolsep}{2.3mm}{
\resizebox{\textwidth}{!}{
\begin{tabular}{c|l|cccc|cccc|cccc}
\toprule[1.2pt]
\textbf{} & \multicolumn{1}{c|}{\textbf{}} & \multicolumn{4}{c|}{\textbf{Games}} & \multicolumn{4}{c|}{\textbf{MicroLens-50K}} & \multicolumn{4}{c}{\textbf{Book}} \\ 
\textbf{} & \multicolumn{1}{c|}{\textbf{}} & \multicolumn{4}{c|}{\textbf{1024-shot ($\bm{r}$=2\%)}} & \multicolumn{4}{c|}{\textbf{1024-shot ($\bm{r}$=2\%)}} & \multicolumn{4}{c}{\textbf{1024-shot ($\bm{r}$=1\%)}} \\ 
\multicolumn{1}{l|}{}  & \textbf{Methods} & \textbf{R@10} & \textbf{R@20} & \textbf{N@10} & \multicolumn{1}{c|}{\textbf{N@20}} & \textbf{R@20} & \textbf{R@50} & \textbf{N@20} & \textbf{N@50} & \textbf{R@20} & \textbf{R@50} & \textbf{N@20} & \textbf{N@50} \\ \midrule \midrule
 & \textbf{TF-DCon} & 0.0102 & 0.0157 & 0.0062 & 0.0078 & {{0.0066}} & {0.0099} & {{0.0027}} & {0.0034} & {{0.0104}} & {0.0144} & {{0.0083}} & {0.0092} \\
 & \textbf{RecRanker} & 0.0112 & 0.0166 & 0.0074 & 0.0090 & 0.0024 & 0.0042 & 0.0011 & 0.0014 & 0.0108 & {0.0145} & {\ul 0.0090} & {\ul 0.0097} \\
 & \textbf{CCS} & {\ul 0.0164} & 0.0246 & 0.0097 & 0.0122 & 0.0096 & 0.0131 & 0.0041 & 0.0049 & {\ul 0.0110} & 0.0145 & 0.0088 & 0.0096 \\
 & \textbf{GraNd} & 0.0158 & 0.0250 & 0.0098 & 0.0125 & 0.0014 & {{0.0032}} & 0.0006 & {{0.0010}} & {0.0102} & {{0.0136}} & {0.0080} & {{0.0087}} \\
			
 & \textbf{EL2N} & 0.0154 & {\ul 0.0256} & 0.0098 & {\ul 0.0128} & 0.0096 & 0.0045 & 0.0041 & 0.0016 & 0.0107 & {\ul 0.0149} & 0.0085 & 0.0094 \\
 & \textbf{Random} & {0.0163} & {0.0241} & {\ul 0.0100} & {0.0122} & {\ul 0.0108} & {\ul 0.0151} & {\ul 0.0044} & {\ul 0.0054} & {0.0099} & {0.0134} & {0.0083} & {0.0090} \\
\multirow{-7}{*}{\textbf{BIGRec}} & \cellcolor{gray!16}\textbf{DEALRec} & \cellcolor{gray!16}\textbf{0.0181*} & \cellcolor{gray!16}\textbf{0.0276*} & \cellcolor{gray!16}\textbf{0.0115*} & \cellcolor{gray!16}\textbf{0.0142*} & \cellcolor{gray!16}\textbf{0.0124*} & \cellcolor{gray!16}\textbf{0.0160*} & \cellcolor{gray!16}\textbf{0.0055*} & \cellcolor{gray!16}\textbf{0.0064*} & \cellcolor{gray!16}\textbf{0.0117*} & \cellcolor{gray!16}\textbf{0.0155*} & \cellcolor{gray!16}\textbf{0.0096*} & \cellcolor{gray!16}\textbf{0.0104*} \\ \midrule
 & \textbf{TF-DCon} & 0.0051 & 0.0074 & 0.0033 & 0.0040 & 0.0006 & 0.0057 & 0.0002 & 0.0013 & 0.0028 & 0.0051 & 0.0020 & 0.0027 \\
 & \textbf{RecRanker} & 0.0028 & 0.0045 & 0.0019 & 0.0024 & {\ul 0.0043} & {\ul 0.0064} & {\ul 0.0011} & {\ul 0.0014} & {0.0027} & {0.0052} & {0.0018} & {0.0025} \\
 & \textbf{CCS} & 0.0050 & 0.0084 & 0.0031 & 0.0041 & 0.0026 & 0.0061 & 0.0010 & 0.0013 & 0.0026 & 0.0048 & 0.0018 & 0.0024 \\
 & \textbf{GraNd} & 0.0042 & 0.0053 & 0.0027 & 0.0030 & 0.0006 & 0.0014 & 0.0003 & 0.0005 & 0.0008 & 0.0020 & 0.0006 & 0.0010 \\
 & \textbf{EL2N} & 0.0034 & 0.0048 & 0.0024 & 0.0029 & 0.0011 & 0.0016 & 0.0004 & 0.0004 & 0.0005 & 0.0015 & 0.0004 & 0.0007 \\
 & \textbf{Random} & {\ul 0.0062} & {\ul 0.0102} & {\ul 0.0039} & {\ul 0.0051} & {0.0037} & {0.0059} & {0.0011} & {0.0014} & {\ul 0.0033} & {\ul 0.0066} & {\ul 0.0022} & {\ul 0.0031} \\
\multirow{-7}{*}{\textbf{TIGER}} & \cellcolor{gray!16}\textbf{DEALRec} & \cellcolor{gray!16}\textbf{0.0074*} & \cellcolor{gray!16}\textbf{0.0114*} & \cellcolor{gray!16}\textbf{0.0062*} & \cellcolor{gray!16}\textbf{0.0074*} & \cellcolor{gray!16}\textbf{0.0058*} & \cellcolor{gray!16}\textbf{0.0076*} & \cellcolor{gray!16}\textbf{0.0020*} & \cellcolor{gray!16}\textbf{0.0020*} & \cellcolor{gray!16}\textbf{0.0039*} & \cellcolor{gray!16}\textbf{0.0076*} & \cellcolor{gray!16}\textbf{0.0026*} & \cellcolor{gray!16}\textbf{0.0037*} \\ \bottomrule[1.2pt]
\end{tabular}
}}
\label{tab:overall_performance}
\end{table*}

\begin{table*}[t]
\setlength{\abovecaptionskip}{0.05cm}
\setlength{\belowcaptionskip}{0.2cm}
\caption{Performance comparison between DEALRec under 1024-shot fine-tuning and the full fine-tuning of the BIGRec in terms of both accuracy and time costs. ``\%Improve.'' denotes the relative improvement achieved by DEALRec compared to the full fine-tuning. Models are trained for 50 epochs with the early stopping strategy.}
\setlength{\tabcolsep}{2.0mm}{
\resizebox{\textwidth}{!}{
\begin{tabular}{l|ccccc|ccccc|ccccc}
\toprule[1.2pt]
 & \multicolumn{5}{c|}{\textbf{Games}} & \multicolumn{5}{c|}{\textbf{MicroLens-50K}} & \multicolumn{5}{c}{\textbf{Book}} \\
 & \textbf{R@10$\uparrow$} & \textbf{R@20$\uparrow$} & \textbf{N@10$\uparrow$} & \textbf{N@20$\uparrow$} & \textbf{Time$\downarrow$} & \textbf{R@20$\uparrow$} & \textbf{R@50$\uparrow$} & \textbf{N@20$\uparrow$} & \textbf{N@50$\uparrow$} & \textbf{Time$\downarrow$} & \textbf{R@20$\uparrow$} & \textbf{R@50$\uparrow$} & \textbf{N@20$\uparrow$} & \textbf{N@50$\uparrow$} & \textbf{Time$\downarrow$} \\ \midrule\midrule
\textbf{Full} & 0.0169 & 0.0233 & 0.0102 & 0.0120 & 36.87h & 0.0081 & 0.0136 & 0.0038 & 0.0053 & 66.64h & 0.0076 & 0.0108 & 0.0060 & 0.0068 & 84.77h \\
\textbf{DEALRec} & 0.0181 & 0.0276 & 0.0115 & 0.0142 & 1.67h & 0.0124 & 0.0160 & 0.0055 & 0.0064 & 1.23h & 0.0117 & 0.0155 & 0.0096 & 0.0104 & 1.93h \\
\cellcolor[HTML]{ECF4FF}\textbf{\% Improve.} & \cellcolor[HTML]{ECF4FF}\textbf{7.10\%} & \cellcolor[HTML]{ECF4FF}\textbf{18.45\%} & \cellcolor[HTML]{ECF4FF}\textbf{12.75\%} & \cellcolor[HTML]{ECF4FF}\textbf{18.33\%} & \cellcolor[HTML]{ECF4FF}\textbf{-95.47\%} & \cellcolor[HTML]{ECF4FF}\textbf{53.09\%} & \cellcolor[HTML]{ECF4FF}\textbf{17.65\%} & \cellcolor[HTML]{ECF4FF}\textbf{44.74\%} & \cellcolor[HTML]{ECF4FF}\textbf{20.75\%} & \cellcolor[HTML]{ECF4FF}\textbf{-98.15\%} & \cellcolor[HTML]{ECF4FF}\textbf{53.95\%} & \cellcolor[HTML]{ECF4FF}\textbf{43.52\%} & \cellcolor[HTML]{ECF4FF}\textbf{60.00\%} & \cellcolor[HTML]{ECF4FF}\textbf{52.94\%} & \cellcolor[HTML]{ECF4FF}\textbf{-97.72\%} \\ \bottomrule[1.2pt]
\end{tabular}
}}
\label{tab:comparison_with_full}
\end{table*}

\subsubsection{\textbf{Implementation.}}\label{sec:implementation}

As for the two backend LLM-based recommender models, we follow the original settings in their paper for implementation. We employ LLaMA-7B for BIGRec and transformer-based architecture for TIGER as in their paper~\cite{rajput2023recommender}. 
All fine-tuning experiments are conducted on four NVIDIA RTX A5000 GPUs. 
Besides, we adopt the parameter-efficient fine-tuning technique LoRA~\cite{hu2021lora} to fine-tune BIGRec and fully fine-tune the parameters of TIGER. 
We utilize SASRec~\cite{kang2018self}, a representative sequential recommender model, as the surrogate model in DEALRec. 
We set the iteration number $T$ for HVP estimation at 5000, and search the regularization strength $\lambda$ in $\{0.1, 0.3, 0.5, 1.0, 2.0\}$. 
For cluster-based methods, the number of clusters $K$ is explored in $\{25,50,75\}$. 
As for the coreset selection methods that require the training of LLMs, we consider a feasible implementation~\cite{coleman2020selection} by executing them on the same surrogate model as DEALRec. 



\subsection{Overall Performance (RQ1)}\label{sec:overall_performance}

The results of the baselines and DEALRec with two competitive backend LLM-based recommender models on three datasets under few-shot fine-tuning (1024 samples) are presented in Table~\ref{tab:overall_performance}, from which we have the following observations: 
\begin{itemize}[leftmargin=*]
    \item All methods with BIGRec typically yield better performance than those with TIGER, which is attributed to two reasons: 1) BIGRec employs a larger LLM (\ie LLaMA-7B) compared to TIGER, thereby benefiting from the stronger generalization ability of large-sized LLMs~\cite{lin2023multi}; and 
    2) BIGRec leverages item titles to present the user sequence, leading to better utilization of world knowledge in LLMs. In contrast, TIGER learns extra item tokens for LLMs. 
    This might result in cold-start item issues since only limited item tokens are learned while others are maintained randomly initialized under the few-shot fine-tuning setting. 
    \item Among all coreset selection baselines, difficulty-based (GraNd, EL2N) methods generally perform better than diversity-based methods (TF-DCon, RecRanker). 
    This is reasonable since diversity-based methods merely heuristically encourage selecting users with divergent preference, which lacks the assessments of their contributions to the model training. 
    In contrast, GraNd and EL2N use pre-defined metrics to measure the sample difficulty and select the samples with larger scores, which encourages selecting the samples that are more informative for models' optimization. 
    Besides, CCS improves EL2N in most cases, as it maintains easy samples for selection, thus compensating the knowledge of recommendation data from high-density areas. 
    
    \item Another interesting observation is that random sampling yields competitive performance or even outperforms other coreset selection methods in some cases, which might attributed to two possible reasons:
    1) Uniformly selected user sequences preserve high coverage of the original training distribution compared to other baselines, which ensures a high probability of guaranteed bound for low empirical risk~\cite{zheng2022coverage}. 
    This observation is also consistent with the findings in~\cite{guo2022deepcore}. 
    2) The inferior performance of some coreset selection methods also might be caused by the implementation settings (Section~\ref{sec:implementation}), where they may suffer from the
    learning ability gap between the surrogate model and LLMs. (\cf Section~\ref{sec:gap_regularization}). 
    

    \item DEALRec significantly outperforms all coreset selection methods across the three datasets. 
    The consistent performance improvements on both backend models validate the superiority of DEALRec in identifying influential samples for LLMs' adaptation to the recommendation data. 
    The superior performance is attributed to: 
    1) the accurate and efficient estimation of the influence on empirical risk, \ie overall performance by removing a sample in training; and 
    2) the gap regularization based on the effort score to penalize the easy samples for LLMs. By emphasizing the non-trivial samples specifically for LLMs, gap regularization alleviates the learning ability gap between the surrogate model and the LLMs. 
\end{itemize}

\vspace{3pt}
\noindent$\bullet\quad$\textbf{Comparison with full fine-tuning.} 
We further compare DEALRec with BIGRec under full training \wrt accuracy and efficiency, as presented in Table~\ref{tab:comparison_with_full}. We can find that: 
1) DEALRec achieves higher performance compared to the model trained by the full data, indicating the effectiveness of DEALRec for high accuracy. 
The inferior performance of BIGRec under full training also implies that not all user sequences are informative for model training, or even harmful to the training, \eg false negative interactions. 
This has also been observed in CTR prediction~\cite{wu2023dataset} and has been discussed in~\cite{agarwal2020contextual} from the view of data redundancy. 
2) DEALRec significantly reduces the time costs for LLMs' fine-tuning ({97.11\%} reduction of fine-tuning costs on average). With the remarkably declined training costs, DEALRec has the potential to facilitate real-world applications of LLM-based recommender models. 

\subsection{In-depth Analysis} 

\subsubsection{\textbf{Ablation Study (RQ2).}}\label{sec:ablation} 

\begin{figure}[t]
\setlength{\abovecaptionskip}{-0.15cm}
\setlength{\belowcaptionskip}{-0cm}
  \centering 
  \hspace{-0.7in}
  \subfigure{
    \includegraphics[height=1.35in]{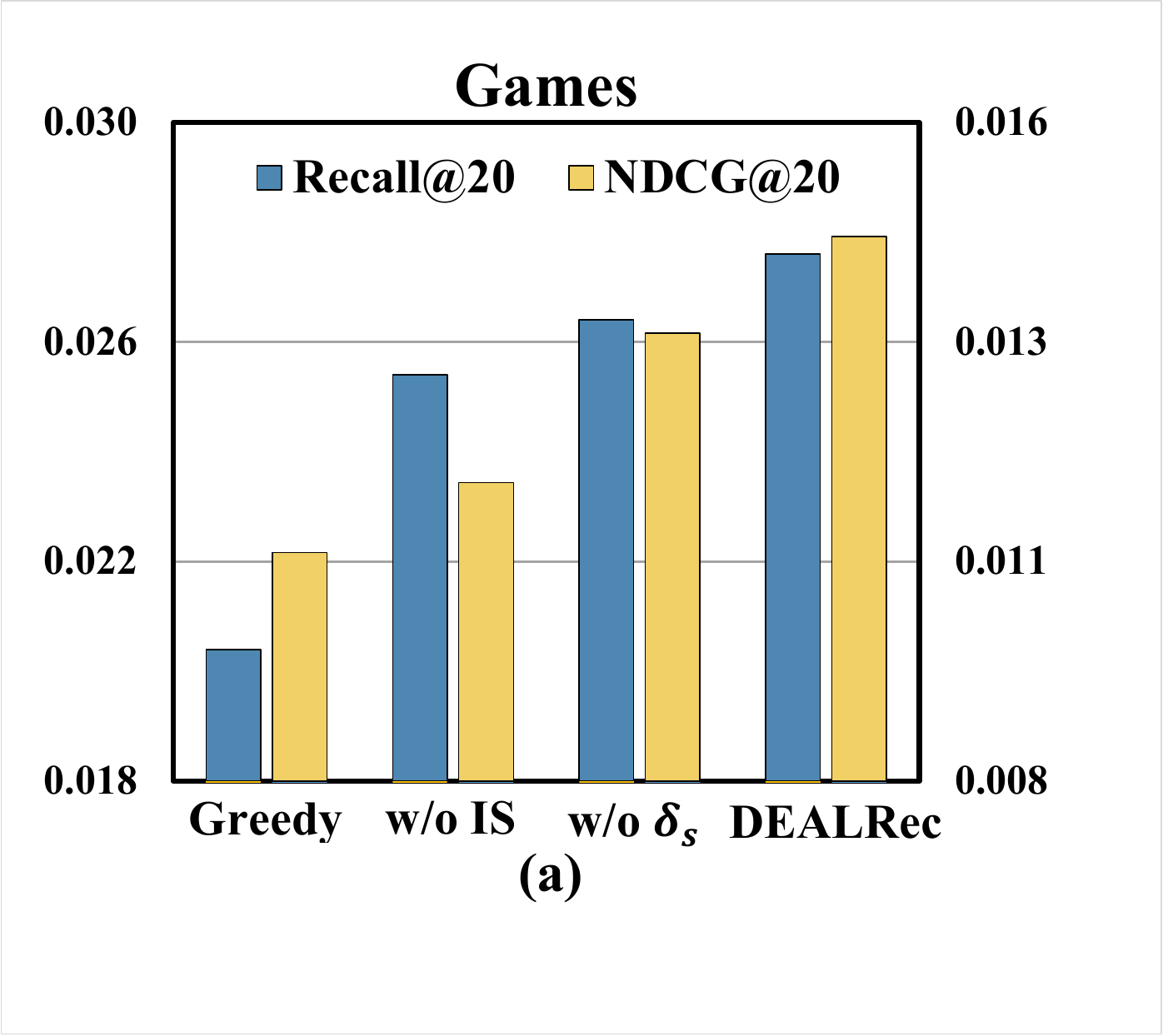}} 
  \hspace{-0.105in}
  \subfigure{
     \includegraphics[height=1.35in]{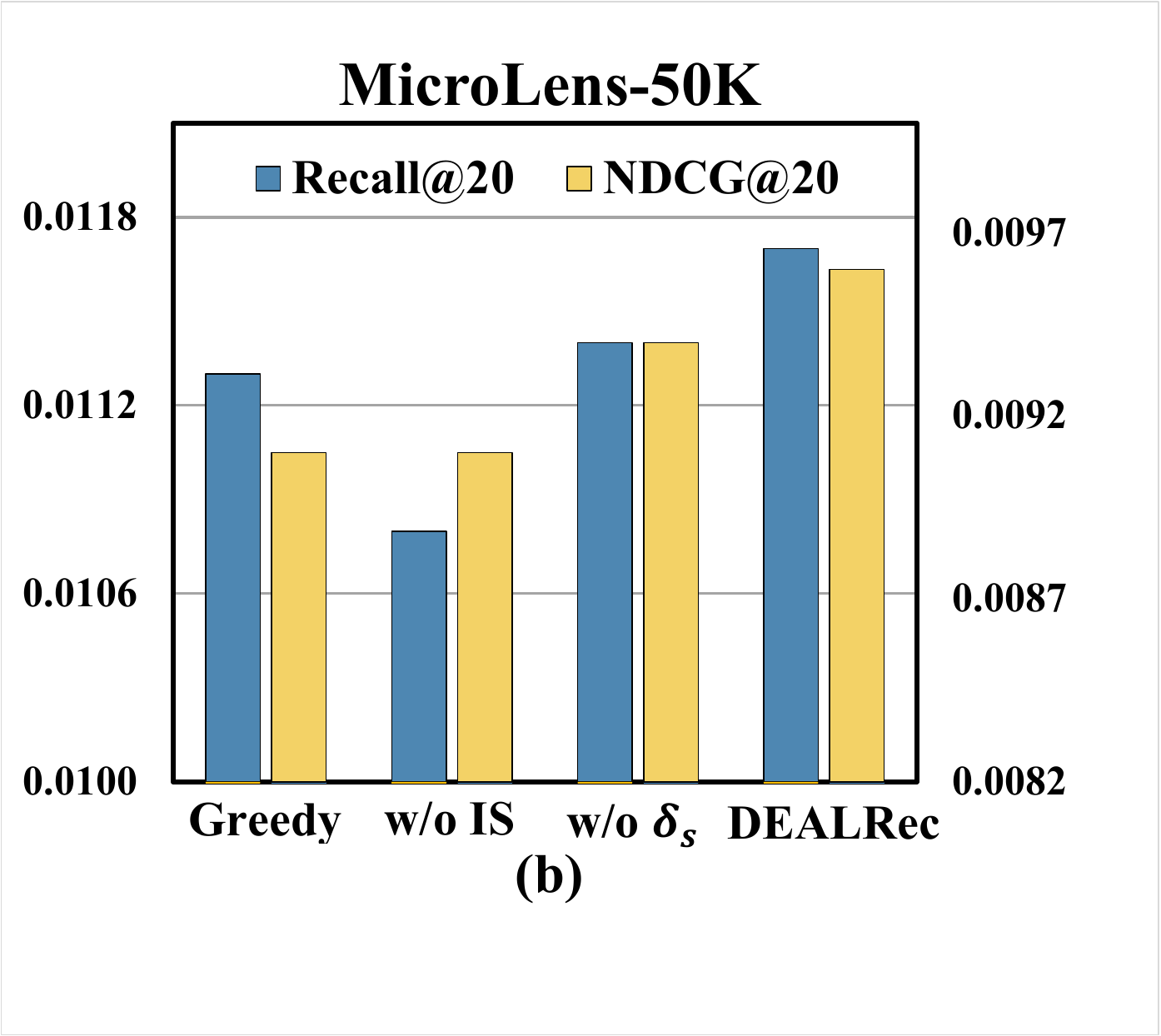}} 
  \hspace{-0.7in} 
\caption{Ablation study of the influence score, effort score, and coverage-enhanced sample selection strategy.}
  \label{fig:ablation}
  \vspace{-0.3cm}
\end{figure}

To study the effectiveness of each component of DEALRec, \ie influence score, effort score, and coverage-enhanced sample selection strategy, we separately remove the Influence Score (IS) and effort score $\delta_s$, referred to as ``w/o IS'' and ``w/o $\delta_s$'', respectively. 
Besides, we replace the coverage-enhanced sample selection strategy by greedily selecting the samples with higher scores, denoted as ``Greedy''. 
From the results presented in Figure~\ref{fig:ablation}, we can observe that: 
removing either the influence score or effort score will cause performance drops. 
This validates the effectiveness of 
1) the assessment of overall performance change caused by removing samples from training; 
2) additional signals of learning ability captured from LLMs as regularization, alleviating the gap between the surrogate model and the LLMs. 
Moreover, simply selecting the samples with higher overall scores might weaken the learning of distinct user behaviors and item knowledge (inferior performance of ``Greedy''), as discussed in Section~\ref{sec:sample_selection}.

\subsubsection{\textbf{Robustness on different surrogate model (RQ2).}} 
\begin{table}[t]
\setlength{\abovecaptionskip}{0.05cm}
\setlength{\belowcaptionskip}{0.2cm}
\caption{Performance comparison between DEALRec with different surrogate models and the BIGRec under full training. ``Time'' presents the time costs for training the surrogate model on a single NVIDIA RTX A5000.}
\setlength{\tabcolsep}{2.6mm}{
\resizebox{0.46\textwidth}{!}{
\begin{tabular}{l|ccccc}
\toprule[1pt]
 & \textbf{R@10$\uparrow$} & \textbf{R@20$\uparrow$} & \textbf{N@10$\uparrow$} & \textbf{N@20$\uparrow$} & \textbf{Time$\downarrow$} \\ \midrule
\textbf{Full} & 0.0169 & 0.0233 & 0.0102 & 0.0120 & / \\
\textbf{BERT4Rec} & 0.0175 & 0.0258 & 0.0103 & 0.0128 & 0.76h \\
\textbf{SASRec} & 0.0181 & 0.0276 & 0.0115 & 0.0142 & 0.45h \\
\textbf{DCRec} & 0.0211 & 0.0283 & 0.0117 & 0.0137 & 0.61h \\ \bottomrule[1pt]
\end{tabular}
}}
\label{tab:surrogate_robustness}
\end{table}

To further assess the generalization ability of DEALRec on different surrogate models, we employ three representative sequential recommender models, \ie BERT4Rec~\cite{sun2019bert4rec}, SASRec~\cite{kang2018self}, and  DCRec~\cite{yang2023debiased} as the surrogate models, respectively. 
From the results in Table~\ref{tab:surrogate_robustness}, 
we can find that: 
1) DEALRec with the three surrogate models consistently outperforms BIGRec under full fine-tuning. 
This demonstrates the strong robustness of DEALRec on different surrogate models. 
2) Different surrogate models cause some fluctuations in accuracy. 
This is reasonable because different model architectures express user behavior and item knowledge differently, 
possibly resulting in varied selected samples which will affect the performance. 
3) SASRec exhibits the least time costs for training and achieves competitive performance among the three surrogate models. 
Therefore, SASRec could be a good choice of surrogate model for DEALRec in real-world deployments.

\subsubsection{\textbf{Effect of selection ratio $\bm{r}$ (RQ3).}}

\begin{figure}[t]
\setlength{\abovecaptionskip}{-0.2cm}
\setlength{\belowcaptionskip}{-0.2cm}
  \centering 
  \hspace{-0.2in}
  \subfigure{
    \includegraphics[height=1.32in]{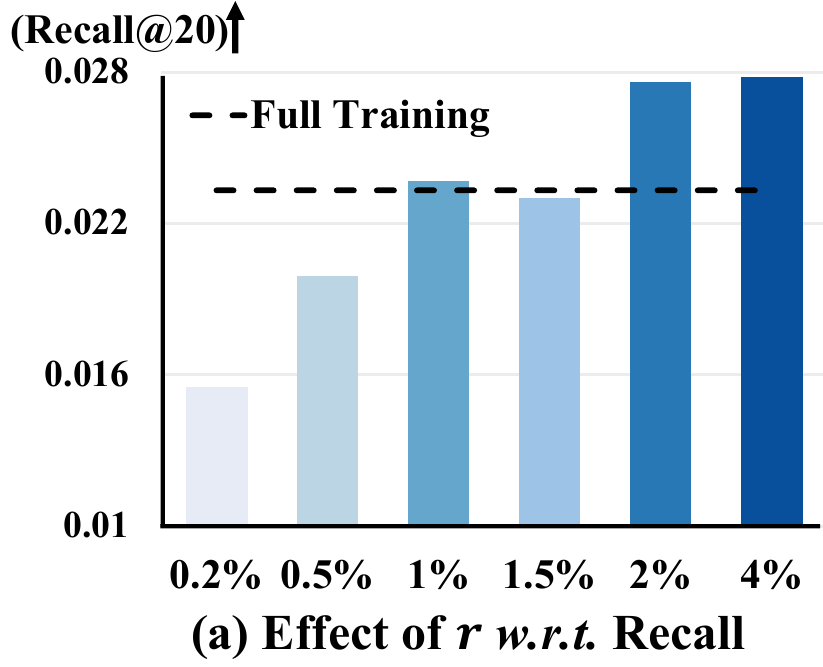}} 
  \hspace{-0.105in}
  \subfigure{
     \includegraphics[height=1.32in]{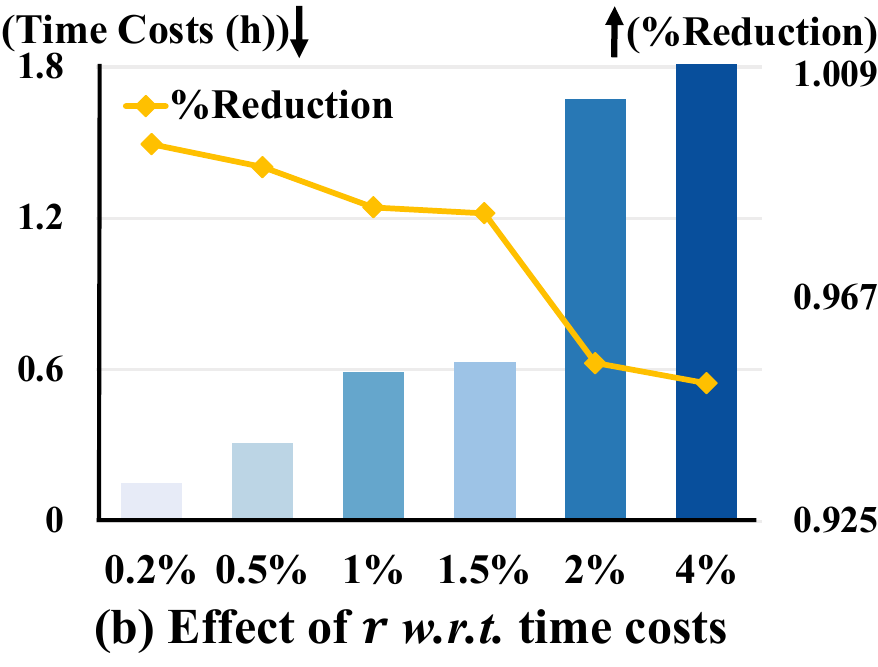}} 
\caption{Performance of DEALRec with different selection ratio $r$ \wrt accuracy and efficiency on Games.}
  \label{fig:n_fewshot}
  \vspace{-0.2cm}
\end{figure}

To investigate the effect of selection ratio $r$ on DEALRec on both accuracy and efficiency, we vary the ratio ${r}$ from $0.2\%$ (128-shot) to $4\%$ (4096-shot) and present the results in Figure~\ref{fig:n_fewshot}. 
It is observed that: 
1) The recommendation accuracy rapidly improves as the number of selected samples increases from $0.2\%$ to $1\%$, surpassing the full training when $r=1\%$. 
Besides, if we continuously increase the selection ratio from $2\%$ to $4\%$, the benefits from additional samples gradually diminish and only minor improvements in accuracy are observed. 
We suspect that the gap between and the recommendation data mainly resides in a small subset of the representative user behaviors, which is what DEALRec aims to identify.
2) Meanwhile, although the time costs for fine-tuning LLMs gradually increase because of additional samples, 
the cost reduction compared to the full training still reaches over 94$\%$. 
3) Empirically, setting $r=1\%$ is recommended to achieve comparable performance to full fine-tuning and low costs.

\subsubsection{\textbf{User group evaluation (RQ3).}} 
\begin{figure}[t]
\setlength{\abovecaptionskip}{-0.1cm}
\setlength{\belowcaptionskip}{-0.1cm}
  \centering 
  \hspace{-0.7in}
  \subfigure{
    \includegraphics[height=1.5in]{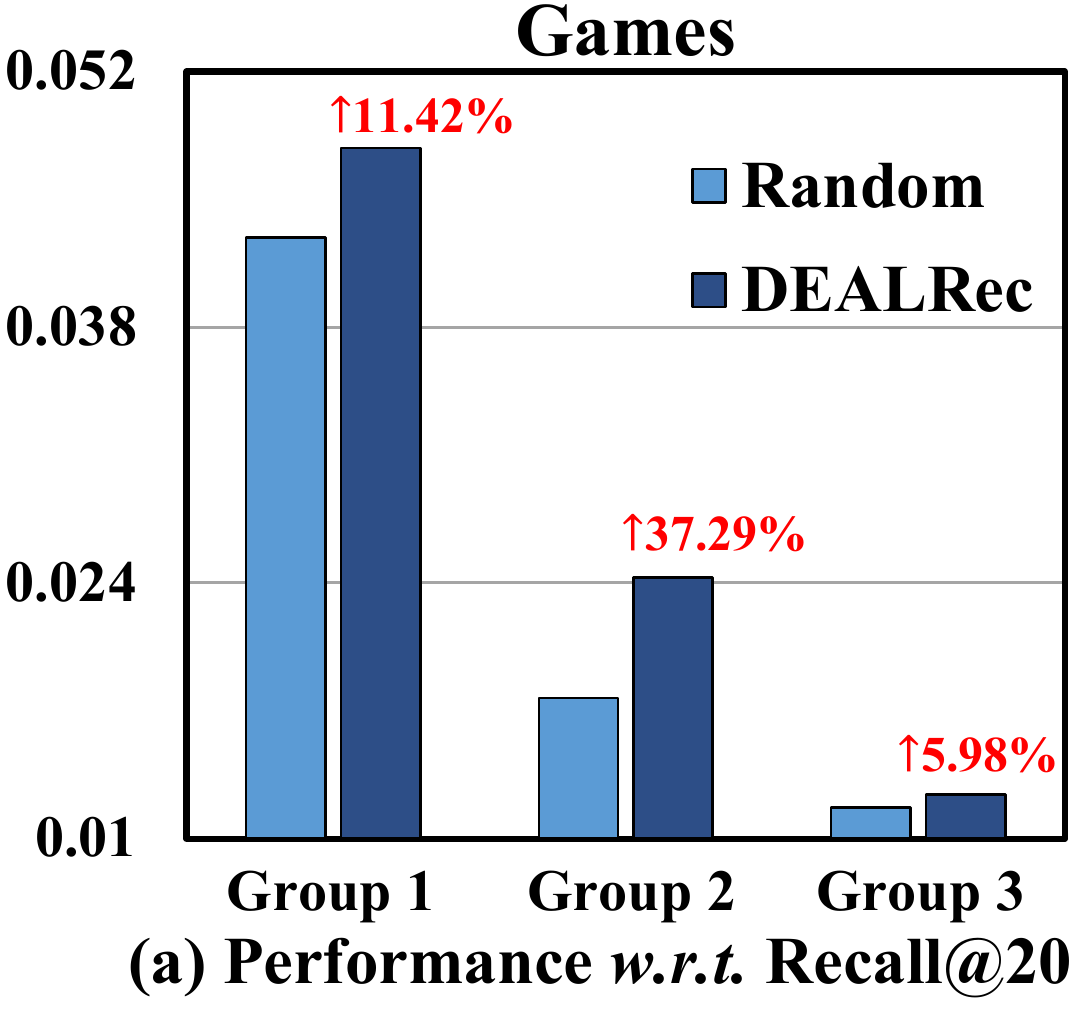}} 
  \subfigure{
     \includegraphics[height=1.5in]{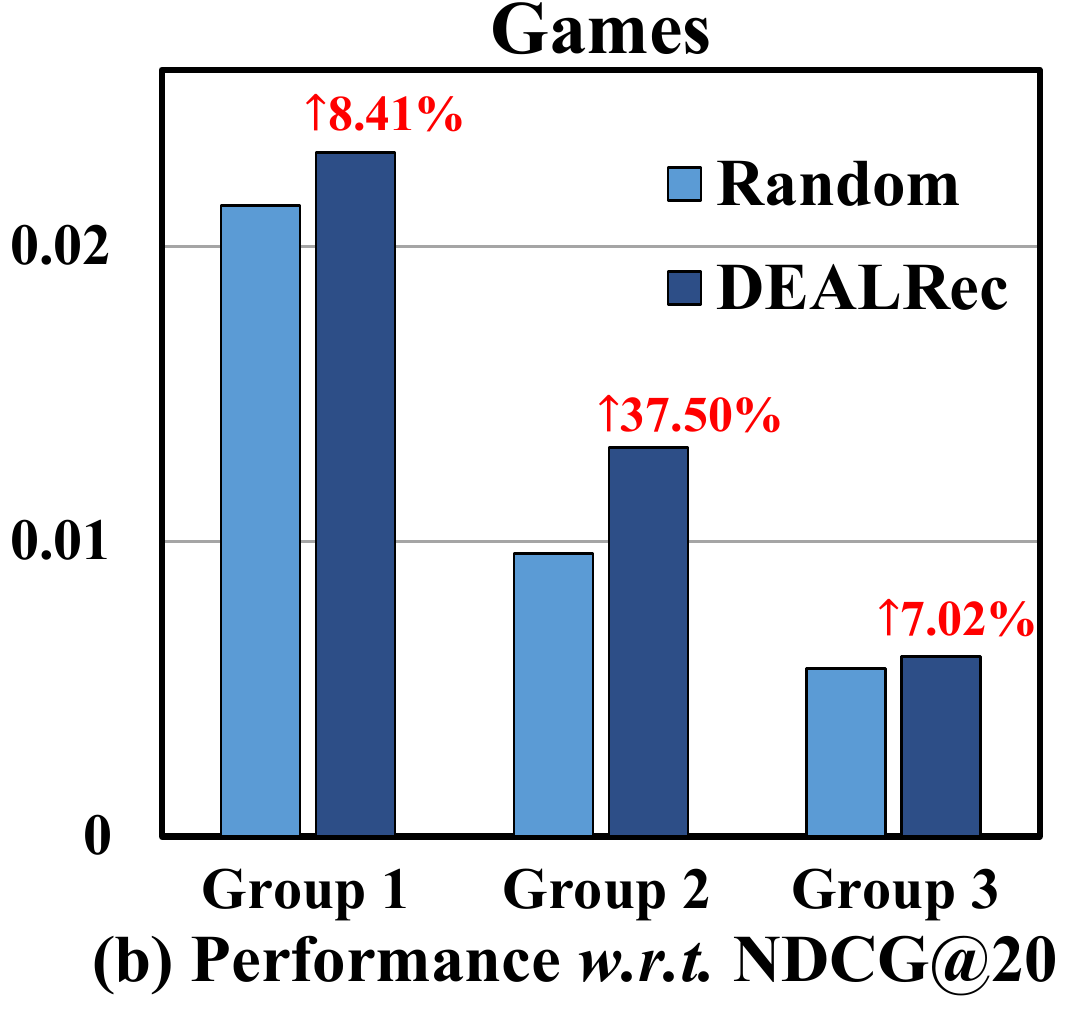}} 
  \hspace{-0.7in} 
\caption{Performance of DEALRec over easy to difficult samples (Group 1 to Group 3).}
  \label{fig:user_group}
  \vspace{-0.2cm}
\end{figure}

To study how DEALRec achieves superior overall performance, we test DEALRec over user sequences of different difficulties. 
Specifically, we calculate the loss of each user sequence via the model trained by randomly selected few-shot samples; 
we then divide the users into three groups according to their loss values, from the easier samples with smaller loss (Group 1) to the harder samples with larger loss (Group 3). 
The results of each group of DEALRec and Random on Games are presented in Figure~\ref{fig:user_group}. 
We can find that 
1) the performance of both DEALRec and Random gradually declines from Group 1 to Group 3, because users with larger loss are more difficult to predict. 
Nevertheless, 
2) DEALRec consistently outperforms Random in each group, which validates the effectiveness of DEALRec in considering the influence on overall performance. 

\subsubsection{\textbf{Effect of regularization strength $\bm{\lambda}$.}} 
\begin{figure}[t]
\setlength{\abovecaptionskip}{-0.2cm}
\setlength{\belowcaptionskip}{-0.2cm}
  \centering 
  \hspace{-0.7in}
  \subfigure{
    \includegraphics[height=1.3in]{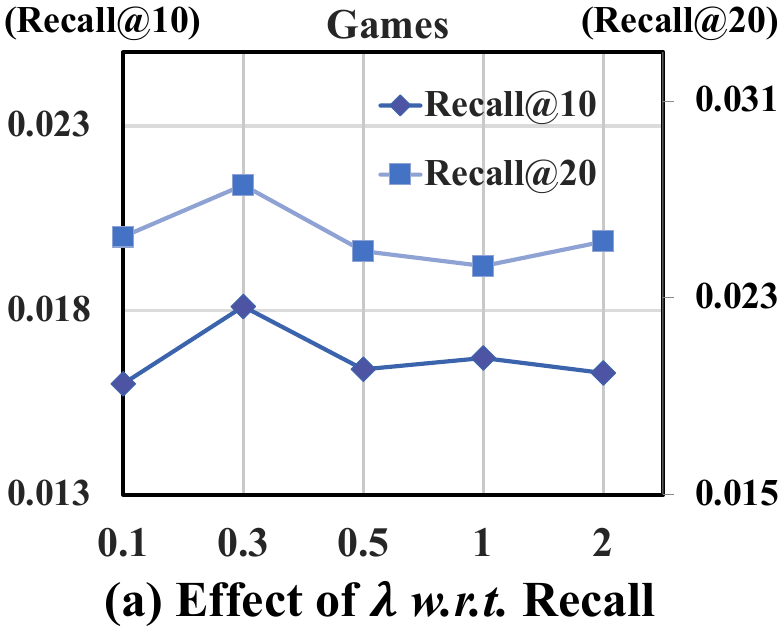}} 
  \subfigure{
     \includegraphics[height=1.3in]{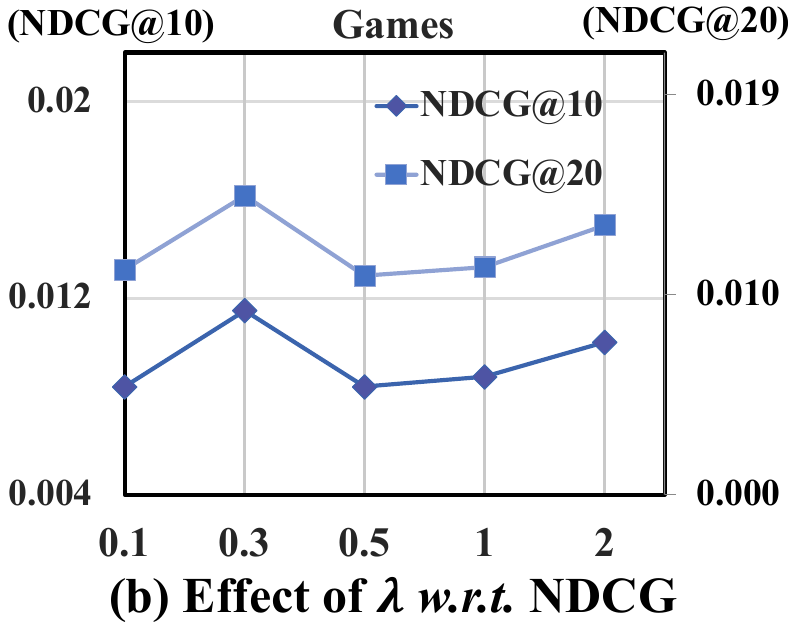}} 
  \hspace{-0.7in} 
\caption{Performance of DEALRec with different $\lambda$.}
  \label{fig:hp_lambda}
  \vspace{-0.2cm}
\end{figure}

We vary $\lambda$ from $0.1$ to $2$ for DEALRec and evaluate the performance as in Figure~\ref{fig:hp_lambda}. 
From the figures, we can find that: 
1) As we incrementally increase the value of $\lambda$, the overall trend of accuracy has been observed to be generally improved. 
This is due to the gap between the surrogate model and LLMs as discussed in Section~\ref{sec:gap_regularization}, emphasizing the necessity to regularize the influence score to be aligned with the learning ability of the LLMs. 
2) However, blindly pursuing larger lambda is not necessarily beneficial. We should carefully balance between the performance-driven influential samples from the surrogate model and the difficult samples for the LLMs. 


%% file: 4_related_work.tex
\section{Related Work}\label{sec:related_work}

\subsection{LLM-based Recommendation}
Leveraging LLMs for recommendation has gained remarkable attention recently~\cite{ren2023representation,zhang2024transfr}, showcasing their potential across various recommendation tasks~\cite{bao2023tallrec,lin2023multi,gong2023unified}.  
Some early studies explore the recommendation ability of powerful LLMs through in-context-learning ability~\cite{dai2023uncovering,sun2023chatgpt}. 
Nevertheless, the performance of LLMs is limited without extra fine-tuning over the domain-specific recommendation data~\cite{bao2023tallrec}. 
To fully leverage the potential of LLMs for recommendation, a series of work studies various fine-tuning strategies tailored for recommendation tasks~\cite{li2023prompt,zhang2023recommendation,gong2023unified,zhang2023lightfr,lv2023duet,lv2024intelligent}. 
However, fine-tuning LLMs requires extensive computational resources and time costs, thus hindering real-world applications. 
Therefore, it is crucial to enhance the fine-tuning efficiency of LLM-based recommender models. 
In this work, we propose the task of data pruning for efficient LLM-based recommendation, aiming to identify representative samples for LLMs' few-shot fine-tuning.

\subsection{Coreset Selection}
Coreset selection has been widely studied in both traditional machine learning and deep learning~\cite{wei2015submodularity,yang2022dataset}, benefiting many downstream tasks such as data-efficient learning~\cite{toneva2018empirical}, neural architecture search~\cite{shim2021core}, and active learning~\cite{sener2017active}. 
It aims to select a small but representative subset from the full data that can lead to comparable model performance. 
Previous work mainly falls into two groups: 
1) Heuristic methods~\cite{coleman2020selection,toneva2018empirical,feldman2020neural} typically assume difficult or diverse samples are informative for model training.
2) Optimization-based methods~\cite{yang2022dataset,killamsetty2021grad,kothawade2022prism} leverages the bi-level or discrete optimization techniques to optimize the data subset that can minimize the empirical risk. 
However, heuristic methods might be suboptimal since they overlook the impact of selected samples on empirical risk. 
And optimization-based methods fail to be applied to LLM-based recommendation due to the cumbersome calculation for complex optimization. 
Furthermore, previous methods usually rely on the training of the model on full data for selection, which is infeasible for LLM-based recommendation (\cf Section~\ref{sec:task_formulation}). 

\vspace{2pt}
\noindent$\bullet\quad$\textbf{Data Condensation}~\cite{zhao2020dataset} 
is another potential solution to achieve data-efficient training. 
However, it is intrinsically different from our proposed task of data pruning. 
While it aims to synthesize a small but informative dataset~\cite{zhao2023dataset}, our task targets to identify existing samples that are representative. 
Besides, previous work mainly works for continuous data, which is inapplicable to LLM-based recommendation~\cite{wu2023dataset}. 
TF-DCon~\cite{wu2023leveraging} is recently proposed for content-based recommendation and we compare it in Section~\ref{sec:overall_performance}. 

%% file: 5_conclusion.tex
\section{Conclusion}\label{sec:conclusion}
In this work, we proposed the task of data pruning for efficient LLM-based recommendation, which aims to identify representative samples tailored for LLMs' few-shot fine-tuning. 
Furthermore, we posited two objectives for this data pruning task: 
1) high accuracy targets to select the samples that can lead to low empirical risk; and 
2) high efficiency strives to consume low costs for the data pruning process. 
To this end, we proposed a novel data pruning method, namely DEALRec, to efficiently identify the influential samples with two scores. 
1) The influence score is formulated to estimate the influence of sample removal on empirical risk, which is extended from the influence function and is accelerated through the symmetric property. 
2) We introduced a small-sized surrogate model to calculate the influence score efficiently and proposed the effort score to bridge the gap between the surrogate model and LLMs. 
Empirical results validate the effectiveness of DEALRec in achieving both high efficiency and high accuracy. 

This work proposes a data pruning task for LLM fine-tuning, opening up a new research direction for efficient LLM-based recommendation and leaving many promising future directions for future work. 
1) It is worthwhile to apply DEALRec to more LLM-based recommender models on more cross-domain datasets, improving fine-tuning performance with limited resources. 
2) Due to the limited context window length of LLMs, it is promising to select the informative interacted items in users' interaction sequences for LLMs' fine-tuning. 
3) Enhancing the inference efficiency of LLM-based recommender models is also a crucial problem for their real-world deployments. 
%